\documentclass[aps,pra,twocolumn,groupedaddress,showpacs]{revtex4}
\usepackage{graphicx,longtable}
\begin{document}


\title{Structure and static response of small silver clusters to an external 
electric field}


\author{M. Pereiro}
\email{fampl@usc.es}
\affiliation{Instituto de Investigaci\'ons Tecnol\'oxicas
and Departamento de F\'{\i}sica Aplicada, Universidade de Santiago 
de Compostela, E-15782, Santiago de Compostela,
Spain}
\author{D. Baldomir}
\affiliation{Instituto de Investigaci\'ons Tecnol\'oxicas
and Departamento de F\'{\i}sica Aplicada, Universidade de Santiago 
de Compostela, E-15782, Santiago de Compostela,
Spain}

\date{\today}

\begin{abstract}
The static response properties and the structural stability of silver clusters in the size range $1\le n \le 23$
have been studied using a linear combination of atomic Gaussian-type orbitals within
the density functional theory in the finite field approach. The Kohn-Sham equations 
have been solved in conjuction with a generalized gradient approximation (GGA) 
exchange-correlation functional.
A proof that the finite basis set GGA calculation holds the Hellmann-Feynman theorem
is also included in the Appendix. The calculated polarizabilities
of silver clusters are compared with the experimental measurements and the jellium model
in the spillout approximation. Despite the fact that the calculated polarizabilities are in 
good agreement with both of them, we have found that the polarizability appears to be 
strongly correlated to the cluster shape and the highest occupied-lowest unoccupied
molecular-orbital gap. 
\end{abstract}

\pacs{36.40.Cg, 32.10.Dk, 31.15.Ew, 61.46.Bc}

\maketitle

\section{Introduction}
Transition-metal clusters play a dominant role 
in cluster physics \cite{heer}. They 
have attracted the interest of many researchers and consequently 
the number of publications on that topic
have experienced a dramatic increase over the past thirty years because metal
clusters exhibit increasingly interesting structural, electronic, catalytic,
as well as optical properties \cite{brack,baletto}. Likewise, 
the biomedical applications 
of nanoclusters have experienced a great impact in the biomedicine community \cite{pankhurst}. 

Among the aforementioned applications and properties of metal clusters, their
optical properties that result from the 
light-matter interaction at the nanoscale level are an area of great 
current interest \cite{kreibig}.
Many of the advances in this area could not have been made possible without the development of
the optical spectroscopy techniques that have been indispensable for elucidating the 
electronic structure of clusters. These experimental techniques can be divided into
two groups, that is, nondestructive and destructive methods. In the former methods,
also called linear response methods, 
a weak electromagnetic field interact with the cluster and it absorbs or scatters light
without undergoing ionization or dissociation whereas in the destructive methods the ionization
is achieved. The nondestructive methods in connection with the linear-response theory
have been extensively used to calculate photoabsorption cross sections and specially 
static dipole polarizabilities. 

The static dipole polarizability is a 
physical observable of metal clusters that has been shown to be closely related to 
the shape and structural geometry. For example,
electronic structure calculations
of small Si clusters show that the polarizability is strongly correlated with the shape
of the clusters \cite{deng}. Likewise, the interplay between theory and experiment is
a powerful tool that serves to identify which cluster is observed in the experiments throughout 
the comparison of the calculated polarizabilities with the experimental ones \cite{moullet}. 
Moreover, the static dipole polarizability is also well-known 
that it is intimately related to 
the shell electronic structure. For example, the noble-metal clusters whose
optical properties have been extensively studied in literature (mainly from the experimental
side) present lower static polarizabilities than alkali metals because they are excellent
examples of spherical shell structure. However, in the case of the silver clusters, the influence of
$d$ electrons in the static polarizability has been less 
studied at least from the theoretical
point of view and it deserves more investigation. This point, among others, is addressed in 
this work.

In this article we have employed a density functional theory-based (DFT) calculation within the
generalized gradient approximation (GGA) to properly account the strong correlation effect of the 
localized $d$ electrons and charge density inhomogeneities. We have studied 
the structural stability and the static response properties of small silver clusters
ranging in size from n=1 up to n=23, where n is the number of atoms forming the cluster. 
The calculated static polarizabilities are only compared with the available experimental
data since that unfortunately, no {\it ab initio} quantum-molecular calculations
have been done so far for the static polarizabilities of silver clusters in the size
range covered by our investigation. The available theoretical data are 
reported in Ref.~\cite{idrobo} for very small silver clusters with 
$1\le n \le 8$. The agreement between our values and those of Idrobo {\it et al.},
who computed the static polarizabilities within the framework of the real 
space finite-different {\it ab initio}
pseudopotential method, is excellent in the case of the GGA approximation, which makes more
valuable the results obtained with the present DFT method.
In addition, we have compared our results with the jellium model in the
spillout approximation and the deviation of our results from this model is explained
in terms of the electronic structure parameters like highest occupied and 
lowest unoccupied molecular orbital (HOMO-LUMO) gaps or structural symmetry.
We also show and discuss how the $d$ electrons affect the static polarizabilities 
since that
when n is sufficiently large, i.e. greater than 18 the polarizability tends to be constant
in contrast with the one of the alkali metals. 
The rest of the paper is organized as follows. In Sec.~\ref{section2}, we
present the theoretical background and the computational details along with the computational
parameters used in this article. Moreover, the structural stability of the silver clusters is studied in detail
in Sec.~\ref{section3}. The results of our calculations and the influence of the 
electronic structure observables in the static polarizabilities are given in Sec.~\ref{section4}.
We conclude with a brief summary of the reported results provided by our 
{\it ab initio} calculations in Sec.~\ref{section5}. In the Appendix we prove that 
the Hellmann-Feynman theorem is satisfied by the finite basis set density 
functional framework in the generalized gradient approximation.

\section{Method and computational details}
\label{section2}
Traditionally there has been two different ways of computing the polarizability 
of molecular systems. Thus, the polarizability is identified either as the
second-order term in the perturbation expansion of the electronic energy
with respect to the applied external uniform electric field or as the linear
response of the dipole moment to that electric field. Both definitions are
equivalent when the Hellmann-Feynman theorem is satisfied. The Hellmann-Feynman
theorem holds for an exact solution of the Schr\"{o}dinger equation and
also for some approximate solutions. In particular, it is satisfied 
by the fully self-consistent finite basis set solutions when the
exchange-correlation (XC) energy is approximated by the
GGA as shown in the Appendix.

We have adopted the dipole moment-based definition 
as our working definition because in this expression the field occurs only
in the first power for the calculation of the polarizability instead of the
second as in the energy expansion. The polarizabilities 
$\alpha_{ij}$ (\{i,j\}=x,y,z or alternatively \{i,j\}=1,2,3) are calculated
by the finite field (FF) method \cite{kurtz} which consists of computing
the electric dipole moment $\mu_i$ of a system under the influence of
an external electric field $F_i$ according to the following finite-difference relation
\begin{equation}
	\alpha_{ij}=\left( \frac{\partial \mu_i(F_j)}{\partial F_j}\right)_{\vec{F}=0}
	=\lim_{F_j\rightarrow 0}\frac{\mu_i(F_j)-\mu_i(-F_j)}{2F_j}.
\end{equation}
In the FF method, one of the most crucial problems to evaluate the derivatives is the choice of 
an appropriate field strength. Several works have assessed the numerical accuracy of polarizability
against different field values and concluded that the best region of linear response
is for field strengths ranging from $10^{-4}$ up to $10^{-2}$ a.u. \cite{vasiliev,zhao}. 
For that reason we have used
a field strength of $5\times 10^{-4}$ a.u. that is applied along the molecular axis.
At least 7 self-consistent field (SCF) runs, with the field strengths  0 and $\pm F_i$, 
are necessary to calculate the polarizability.
Once the polarizability tensor components are computed, the mean static polarizability is 
calculated as $\bar{\alpha}=(\sum{_{i=1}^3}\alpha_{ii})/3$ and the polarizability anisotropy
is defined as
\begin{equation}
	\label{eq:anisotropy}
\Delta\alpha=\sqrt{\frac{\sum\limits _{i,j=1,2\atop i<j}^{2,3}(
\alpha_{ii}-\alpha_{jj})^2+6\sum\limits _{i,j=1,2\atop i<j}^{2,3}\alpha_{ij}^2}{2}} 
\end{equation}
in the general axis frame or without the second addended ($6\sum_{i,j=1,2; 
i<j}^{2,3}\alpha_{ij}^2$) in the coordinate system which makes the second-rank
polarizability tensor $\mathsf{\alpha}$ diagonal. 

\begin{figure*}
\includegraphics[width=17.8cm,angle=0]{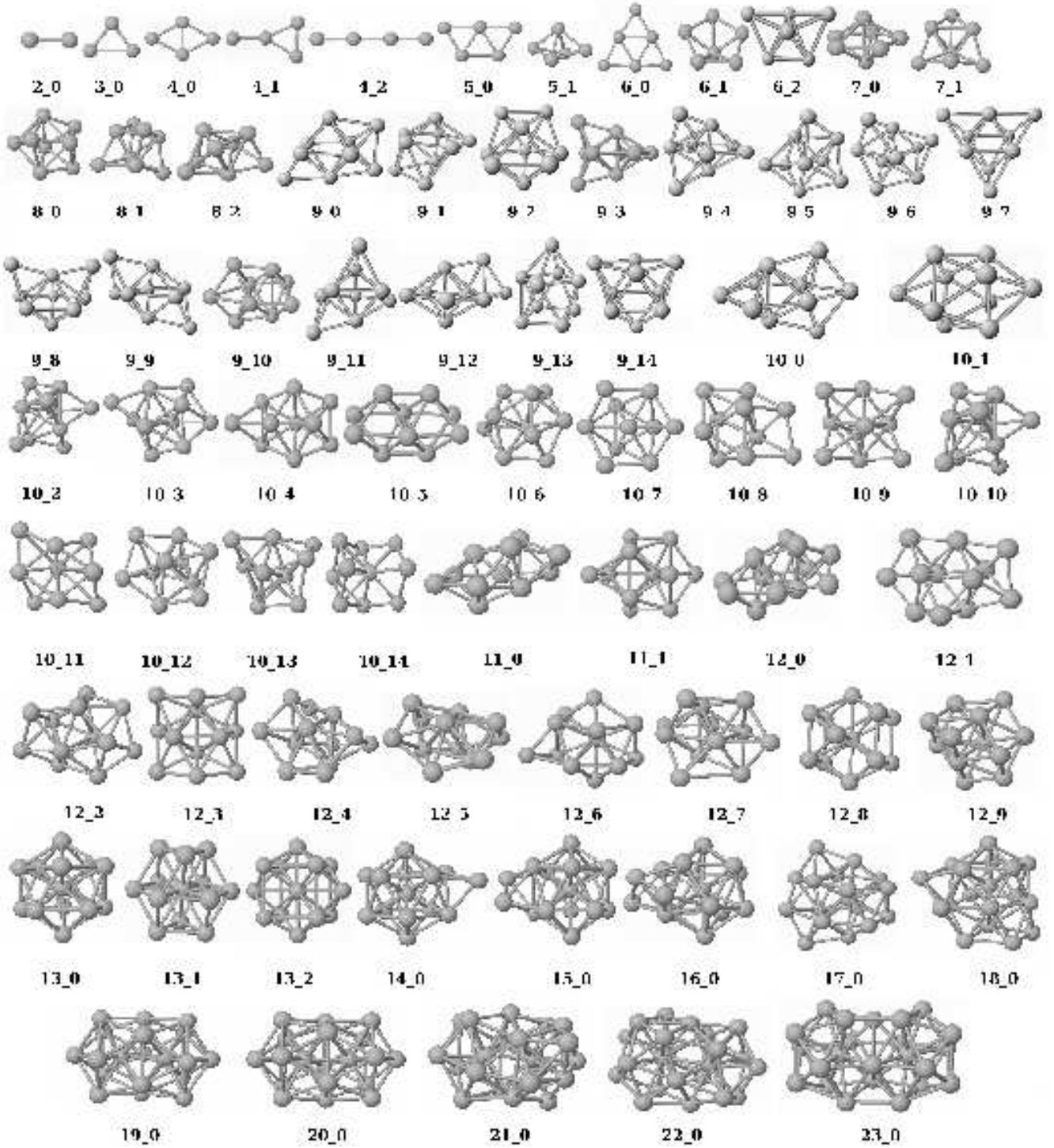}
\caption{\label{fig1}Lowest-energy structures and isomers of Ag$_n$, $n=2-23$, ordered (from left to 
right and top to bottom) by increased size and energy.  The cluster $n$\_$m$ is the $m$th energetic isomer
with $n$ atoms.}
\end{figure*}

With the aim of studying the static response properties of small silver clusters,
Ag$_n$ (2$\le$n$\le$23), we have
performed density functional theory-based calculations consisting of 
a linear combination of Gaussian-type-orbitals (LCGTO) Kohn-Sham density-functional
methodology as it is implemented in  {\sc demon-ks3p5} program \cite{salahub}. 
All-electron spin-unrestricted calculations were carried out
at the GGA level to 
take the XC effects into account \cite{perdew}. 
Local-density approximation (LDA)
sometimes yields inaccurate bond lengths and total energies
due to the insufficiency in describing the strong correlation effects of the
localized $d$ electrons and charge density inhomogeneities. In these regards, the
GGA should be a choice better than LDA \cite{pereiro}.
For this reason, at the beginning of this work and to satisfy ourselves that the 
numerical procedure is reliable, we initiate a search of the functional that better
fitted the calculated bond length of the silver dimer to the experimental one.
The functional developed by
Perdew and Wang \cite{perdew} gave us a bond length of 2.534~\AA, that is in excellent 
agreement with the experimental measure (2.53350~\AA) reported in Ref.~\cite{simard}.
An orbital basis set of contraction pattern (633321/53211*/531+) was used in conjunction
with the corresponding (5,5;5,5) auxiliary basis set for 
describing the $s$, $p$ and $d$ orbitals \cite{huzinaga}. In {\sc demon-ks3p5}, the
electron density is expanded in auxiliary basis functions which are introduced to
avoid the calculation of the $N^4$ scaling Coulomb repulsion energy, where N is the
number of the basis functions. The grid for numerical
evaluation of the XC terms had 128 radial shells of points and each shell had
26 angular points. Spurious one-center contributions to the XC forces, typically 
found in systems with metal-metal bonds when
using a nonlocal functional, are 
eliminated in a similar way as has been done in Ref.~\cite{versluis}.
A wide set of spin multiplicities ranging from 1 to 11 was checked to
ensure that the lowest-energy electronic configuration is reached.
The geometries were fully optimized without symmetry and
geometry constraints using the Broyden-Fletcher-Goldfarb-Shanno
algorithm \cite{broyden}. During the optimization, the convergence criterion 
for the norm of the energy gradient was fixed to $10^{-4}$ a.u. while it was $10^{-7}$ a.u. for 
the energy and $10^{-6}$ a.u. for the charge density. The ground state structures and some of 
the lowest-energy isomers of the silver clusters studied in this article
are illustrated in Fig.~\ref{fig1}, since that the polarizability is closely related to the
geometrical shape of the cluster. 

\section{Structures of silver clusters}
\label{section3}
Figure~\ref{fig1} and Table~\ref{table1} show the structures, relative energies, 
first-neighbor distances and vertical ionization potentials (IP) of the lowest-energy isomers predicted
by the simulations for Ag$_n$ clusters with $2\le n \le 23$. The IPs have been calculated as the
energy difference between the neutral cluster and the cation. We assign labels to clusters such as $n$\_$m$,
where $n$ indicates the number of atoms and the second number gives the rank in increasing energy order.
A huge sampling of trial geometries taken
from the literature was evaluated \cite{rene}. While for these small clusters, it is nearly impossible to search for
all possible geometries, the detailed search that we have carried out gives us some confidence that the 
ground state structures have been found. 

For clusters with size varying from 2 up to 13, we optimized geometries
and calculated energies for 78 isomers. We only reported the lowest-energy structures here, but their 
Cartesian coordinates in Bohr units are available upon request to the authors. We focus on isomers located
50 KJ/mol ($\approx$ 0.5 eV) or less above the lowest-energy structure because this value is considered
to represent an error bar for relative energies computed with GGA functionals \cite{chretien}. Whenever is 
possible we compare the calculated IPs with the experimental ones to rule out the possible candidates to the
ground state geometry. We will focus our attention mainly on the lowest energy cluster geometries which
are different form the cluster structures reported in previous experimental or theoretical studies.
For the rest of clusters, we simply just describe the lowest energy
structures. In Table~\ref{table1}, other low energy isomers are also listed for comparison. 

For the 
clusters with size varying from 14 up to 23, we took the initial guess structures from Ref.~\cite{doye}. 
Moreover, in the 
early stage of the geometry optimization process and with the aim of speeding up the calculations, the
structure of the silver clusters was first optimized in conjunction with a 17-electron scalar relativistic
model core potential designed for the adequate description of the silver dimer bond length \cite{andzelm}.
Once the geometry of the cluster was converged for the model core potential, an all-electron 
structural-relaxation calculation was performed leading to the current lowest-energy structures showed in Fig.~\ref{fig1}. 

For the silver trimer, we have studied only one structure (isosceles triangle) because
the minimum structure is 
well established in the literature \cite{rene,luo}. 
In the case of Ag$_4$, the lowest energy structure that we have found is the planar rhombus.
The lowest-energy structures for Ag$_3$ and Ag$_4$ predicted by our calculations 
are in good agreement with other {\it ab initio} results \cite{rene,luo,bonacic}. 

The ground state structure found for the pentamer is a trapezoid with C$_{2v}$ symmetry.
In literature there is a controversy about the assignment of the structural minimum.
Some theoretical studies predict as the ground state the planar structure \cite{rene,bonacic,liu} 
while other ones predict
the trigonal bipyramid \cite{luo}. On the experimental side, it is worthwhile to mention that different
experimental ESR and Raman spectra of Ag$_5$ have been interpreted by both trigonal bipyramid \cite{howard} and planar
geometries \cite{hasiett}, respectively. As shown in Table~\ref{table1}, the vertical IP calculated for 
the trapezoid
agrees quite well with the experimental result \cite{jackschath} and it is what underpins our predicted 
lowest-energy structure. 

Four structures have been optimized for the Ag$_6$ cluster \cite{footnote1}.
The planar trapezoidal D$_{3h}$ structure resulted to be the most stable
but only 0.079 eV lower in energy than the pentagonal pyramid. Both structures 
exhibit IPs very close to the experimental one and therefore, our ab-initio calculations 
do not allow prediction of only one geometry as the structural minimum.

In the case of the silver heptamer, the pentagonal bipyramid (D$_{5h}$ symmetry) is 
predicted as the lowest energy 
structure while the tricapped tetrahedron is 0.17 eV higher in energy. Although the relative energy and the 
ionization potentials do not allow a clear prediction about the minimum geometry, however most of the
first principles calculations prior to this article have also obtained the pentagonal bipyramid as the
fundamental structure \cite{luo, rene, idrobo, bonacic, bonacic1}.

For the silver octamer, we have decided to optimize as a good candidate to the structural minimum 
the following isomers: a D$_{2d}$ dodecahedron, which can be viewed as a distorted bicapped
octahedron, a T$_d$ tetracapped tetrahedron and a C$_s$ pentagonal bipyramid. Our calculations stabilize
first the dodecahedron, secondly but very close in energy the T$_d$ structure and finally the pentagonal
bipyramid with $\Delta E_{DFT}$=0.181 eV. The comparison between the calculated IPs and the experimental
measurements reported in Ref.~\cite{jackschath} favor more the T$_d$ structure than the D$_{2d}$ one,
however the assignment of the structural minimum is clearly inverted if we use the experimental IP (6.40 eV) 
measured in Ref.~\cite{laihing}. Our predicted minimum structure is also supported by a very recent
reference where the authors determine the lowest-energy structure of Ag$_8$ by comparison of the optical
spectra provided by time-dependent DFT with the experimental findings \cite{pereiro1}.

Fifteen structures have been optimized in the case of the Ag$_9$ cluster. A tricapped-distorted 
octahedron (C$_s$) was
found for the lowest-energy structure which is in good agreement with the reported structure in 
Ref.~\cite{rene}, while in Refs.~\cite{luo,bonacic} the ground state geometry is the bicapped pentagonal bipyramid.
To satisfy ourself that the bicapped pentagonal bipyramid is not the structural minimum, we studied a wide range of
pentagonal bipyramid structures capped with two atoms in different positions. After the geometry optimization, the
final geometry is slightly distorted in most of the cases. The structures are plotted in Fig.~\ref{fig1} and denoted
as 9\_1, 9\_2, 9\_3, 9\_4, 9\_5, 9\_6, 9\_8, 9\_9, 9\_10, 9\_11, and 9\_12. It is worthwhile to mention that from 9\_0 up to
9\_6, the structures are very close in energy and only the IPs of the pentagonal bipyramid structures agrees quite
well with the experimental ones, so in gas-phase experiments is likely to obtain the bicapped pentagonal
bipyramid as the lowest-energy structure. 

Contrary to what happened in the case of nanomers, the lowest-energy geometry of Ag$_{10}$ cluster is a pentagonal
bipyramid-shaped structure, namely, a tricapped pentagonal bipyramid. With the aim of testing the validity of this
structure, we have also optimized a very distorted pentagonal bipyramid but it finally converged to the 10\_0 structure.  
The second structure is a D$_{4d}$ bicapped square antiprism and it is only 0.080 eV higher in energy with respect to the 
ground state. This structure has been predicted as the fundamental one in Ref.~\cite{luo}, nevertheless the comparison
of the calculated IPs reported in Table~\ref{table1} with the experimental value clearly show that the 10\_0 structure
is the best candidate to be the minimum structure. 

The guess structures for Ag$_{11}$ and Ag$_{12}$ were obtained by adding atoms to the pentagonal bipyramid structure or 
by removing them from a 13-atom O$_h$ cuboctahedron or from the icosahedral packing. Thus, 
the lowest-energy structures come from the pentagonal bipyramid shape in both cases while the structures coming from
the cuboctahedron or the icosahedral packing are less favored energetically.
In the case of the Ag$_{11}$ geometries, we have studied 9 isomers but except for the 
11\_1 structure, the other ones are energetically far from the lowest-energy structure by an amount greater than 0.5 eV.
While the relative energies clearly show that the fundamental structure is the one labeled with 11\_0 
($\Delta E_{DFT}^{11\_0\rightarrow 11\_1}=0.268$ eV), however
the calculated IP ($\Delta$ IP$^{11\_1}$=0.02 eV) predicts the 11\_1 as the minimum structure, and 
consequently these two structures would be probably observed in
experiments. The same situation occurs for Ag$_{12}$ but in this case more structures can be observed in low-temperature
experiments.

Three structures have been selected as possible candidates for the structural minimum of the thirteen-atom silver cluster.
They are the icosahedral geometry (I$_h$ symmetry), the cuboctahedron cluster (O$_h$ symmetry), and the D$_{2h}$ structure
which is a compact portion of the bcc crystal lattice capped with four atoms as it is illustrated in Fig.~\ref{fig1}.
After the relaxation of the structures, the ground state geometry predicted by our calculations is 
the icosahedral structure.
Although the calculated IPs are in general in a relatively good agreement with the experimental ones, 
however they do not add too much information to the determination of the lowest-energy structure. The relative 
energies of the isomers collected in Table~\ref{table1} together with the fact that most of the articles devoted
to the study of the structural properties of small silver clusters predict the icosahedral structure as fundamental one, give
us some confidence that the icosahedral packing is valid for the Ag$_{13}$ cluster \cite{luo,michaelian,erkoc}.

As we comment above, for the rest of the clusters, that is, from Ag$_{14}$ up 
to Ag$_{23}$ we took the initial guess structures 
from Ref.~\cite{doye}. They follow an icosahedral growth sequence capped with a variable number of atoms depending on the
cluster size. 
In general, the calculated IPs are in good agreement with the experimental measurements which make
more valuable the geometries optimized with the {\sc demon-ks3p5} program using as starting point the structures predicted 
in Ref.~\cite{doye}.

\begin{table*}
\caption{\label{table1}Average first-neighbor distance and relative energy of Ag$_n$ cluster isomers with 
$2\le n\le 23$. The vertical ionization potential is compared with the data from Ref.~\cite{jackschath}. The
geometry notation is that of Fig.~\ref{fig1}.
} 
\begin{ruledtabular}  
	\begin{tabular}{lccccccclccccc}
		\multicolumn{14}{c}{ Ag$_2 \longmapsto$ Ag$_{10\_4}$ \hspace{7cm} Ag$_{10\_5} \longmapsto$ Ag$_{23\_0}$} \\ \cline{1-6} \cline{9-14}
        &	&  & & Vertical IP & & & & & & & & Vertical IP & \\ \cline{4-6}\cline{12-14} 
Cluster & d & $\Delta E_{DFT} $ & Calc. & Exp. & $|\Delta$ IP$|$ & & & Cluster & d  & $\Delta E_{DFT} $ & Calc. & Exp. & $|\Delta$ IP$|$ \\
        & (\AA) & (eV)  &  (eV)         & (eV) &    (eV)         & & &         &  (\AA)  &   (eV)                 & (eV)  & (eV) &   (eV)\\
\hline
2\_0 & 2.53 & 0.000 & 7.73 & 7.60 & 0.13 & & &  10\_5 & 2.78 & 0.397 & 6.63 & 6.25 & 0.38\\
3\_0 & 2.79 & 0.000 & 5.67 & 6.20 & 0.53 & & & 10\_6 & 2.81 & 0.495 & 6.26 & 6.25 & 0.01\\
4\_0 & 2.71 & 0.000 & 6.54 & 6.65 & 0.11 & & & 10\_7 & 2.81 & 0.506 & 6.24 & 6.25 & 0.01\\
 4\_1 & 2.63 & 0.242 & 6.51 & 6.65 & 0.14 & & & 10\_8 & 2.78 & 0.553 & 6.36 & 6.25 & 0.11\\
 4\_2 & 2.56 & 0.563 & 6.93 & 6.65 & 0.28 & & & 10\_9 & 2.77 & 0.617 & 5.99 & 6.25 & 0.26\\
 5\_0 & 2.72 & 0.000 & 6.33 & 6.35 & 0.02 & & & 10\_10& 2.78 & 0.626 & 6.04 & 6.25 & 0.21\\
 5\_1 & 2.77 & 0.430 & 6.16 & 6.35 & 0.19 & & & 10\_11& 2.74 & 0.703 & 6.34 & 6.25 & 0.09\\
 6\_0 & 2.72 & 0.000 & 7.22 & 7.15 & 0.07 & & & 10\_12& 2.80 & 0.720 & 6.00 & 6.25 & 0.25\\
 6\_1 & 2.72 & 0.079 & 7.09 & 7.15 & 0.06 & & & 10\_13& 2.79 & 1.151 & 6.14 & 6.25 & 0.11\\
 6\_2 & 2.80 & 0.534 & 6.67 & 7.15 & 0.48 & & & 10\_14& 2.75 & 1.133 & 5.90 & 6.25 & 0.35\\
 7\_0 & 2.79 & 0.000 & 6.36 & 6.40 & 0.04 & & & 11\_0 & 2.81 & 0.000 & 6.66 & 6.30 & 0.36\\
 7\_1 & 2.76 & 0.170 & 6.44 & 6.40 & 0.04 & & & 11\_1 & 2.81 & 0.268 & 6.32 & 6.30 & 0.02\\
 8\_0 & 2.80 & 0.000 & 6.44 & 7.10 & 0.66 & & & 12\_0 & 2.82 & 0.000 & 7.01 & 6.50 & 0.51\\
 8\_1 & 2.76 & 0.006 & 7.33 & 7.10 & 0.23 & & & 12\_1 & 2.79 & 0.117 & 6.91 & 6.50 & 0.41\\
 8\_2 & 2.78 & 0.181 & 6.66 & 7.10 & 0.44 & & & 12\_2 & 2.80 & 0.161 & 6.72 & 6.50 & 0.22\\
 9\_0 & 2.80 & 0.000 & 6.53 & 6.00 & 0.53 & & & 12\_3 & 2.81 & 0.411 & 6.65 & 6.50 & 0.15\\
 9\_1 & 2.81 & 0.039 & 5.83 & 6.00 & 0.17 & & & 12\_4 & 2.81 & 0.417 & 6.50 & 6.50 & 0.00\\
 9\_2 & 2.81 & 0.039 & 5.97 & 6.00 & 0.03 & & & 12\_5 & 2.81 & 0.422 & 6.49 & 6.50 & 0.01\\
 9\_3 & 2.78 & 0.052 & 5.72 & 6.00 & 0.28 & & & 12\_6 & 2.81 & 0.652 & 6.42 & 6.50 & 0.08\\
 9\_4 & 2.78 & 0.056 & 5.83 & 6.00 & 0.17 & & & 12\_7 & 2.80 & 0.802 & 6.36 & 6.50 & 0.14\\
 9\_5 & 2.78 & 0.070 & 5.91 & 6.00 & 0.09 & & & 12\_8 & 2.80 & 1.039 & 6.58 & 6.50 & 0.08\\
 9\_6 & 2.77 & 0.089 & 5.99 & 6.00 & 0.01 & & & 12\_9 & 2.71 & 1.039 & 6.53 & 6.50 & 0.03\\
 9\_7 & 2.78 & 0.145 & 5.77 & 6.00 & 0.23 & & & 13\_0 & 2.87 & 0.000 & 5.75 & 6.34 & 0.59\\
 9\_8 & 2.79 & 0.154 & 5.70 & 6.00 & 0.30 & & & 13\_1 & 2.72 & 0.305 & 6.23 & 6.34 & 0.11\\
 9\_9 & 2.75 & 0.182 & 6.45 & 6.00 & 0.45 & & & 13\_2 & 2.62 & 3.568 & 5.88 & 6.34 & 0.46\\
 9\_10& 2.77 & 0.204 & 6.31 & 6.00 & 0.31 & & & 14\_0 & 2.83 & 0.000 & 5.87 & 6.73 & 0.86\\
 9\_11& 2.79 & 0.211 & 5.86 & 6.00 & 0.14 & & & 15\_0 & 2.83 & 0.000 & 5.82 & 6.40 & 0.58\\
 9\_12& 2.78 & 0.214 & 6.55 & 6.00 & 0.55 & & & 16\_0 & 2.82 & 0.000 & 5.79 & 6.57 & 0.78\\
 9\_13& 2.77 & 0.218 & 6.11 & 6.00 & 0.11 & & & 17\_0 & 2.82 & 0.000 & 5.84 & 6.45 & 0.61\\
 9\_14& 2.77 & 0.221 & 5.91 & 6.00 & 0.09 & & & 18\_0 & 2.82 & 0.000 & 5.95 & 6.53 & 0.58\\
10\_0 & 2.82 & 0.000 & 6.58 & 6.25 & 0.33 & & & 19\_0 & 2.85 & 0.000 & 5.35 & 6.20 & 0.85\\
 10\_1 & 2.77 & 0.080 & 7.27 & 6.25 & 1.02 & & & 20\_0 & 2.85 & 0.000 & 5.39 & 6.45 & 1.06 \\
 10\_2 & 2.78 & 0.158 & 6.68 & 6.25 & 0.43 & & & 21\_0 & 2.84 & 0.000 & 5.36 & 5.90 & 0.54 \\
 10\_3 & 2.79 & 0.205 & 6.51 & 6.25 & 0.26 & & & 22\_0 & 2.82 & 0.000 & 5.56 & 6.04 & 0.48\\
 10\_4 & 2.78 & 0.221 & 6.77 & 6.25 & 0.52 & & & 23\_0 & 2.84 & 0.000 & 5.42 & 6.03 & 0.61\\
 
\end{tabular}  
\end{ruledtabular}  
\end{table*}

\section{Results and discussion}
\label{section4}
\begin{figure}
\includegraphics[width=8.6cm,angle=0]{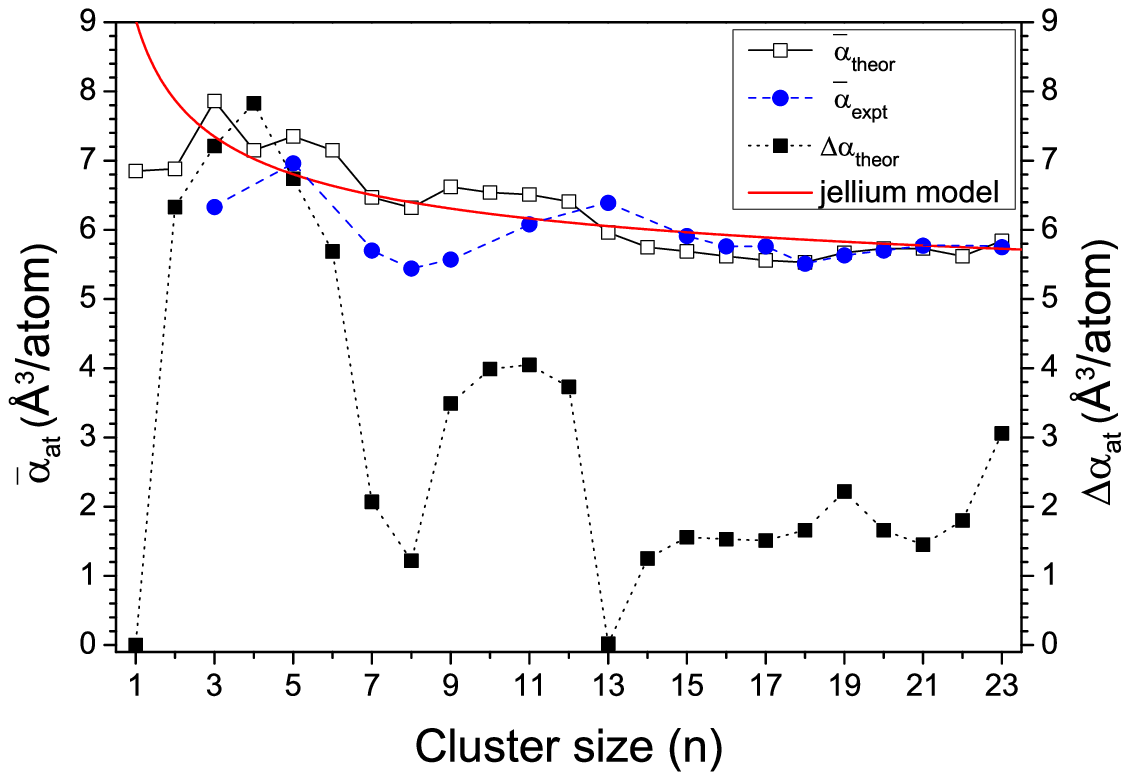}
\caption{\label{fig2}(Color online). Mean static polarizabilities per atom (open squares) and 
the polarizability anisotropies per atom (solid squares)
of Ag$_n$ clusters calculated with the FF method as a function of the cluster size.
The solid circles represent the experimental measurements of the polarizabilities per atom 
taken from Ref.~\cite{fedrigo}. The solid line represents the prediction from the
jellium model. The fitted parameters are given in the text. 
}
\end{figure}

\begin{figure}
\includegraphics[width=8.6cm,angle=0]{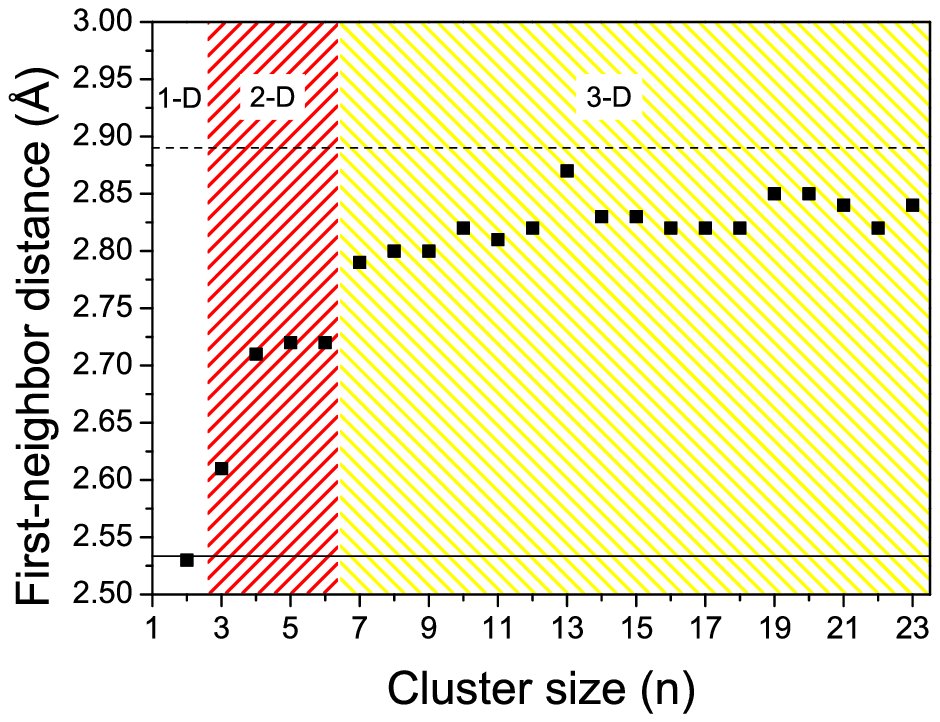}
\caption{\label{fig3}(Color online). First-neighbor distance plotted against the cluster size. The solid line represents
the experimental silver dimer bond length and the dashed line symbolizes the experimental value of the equilibrium 
interatomic distance of fcc silver bulk. The three different areas
stand for the one-dimensional (1-D), bidimensional (2-D) and three-dimensional (3-D) lowest-energy structures of small silver cluster, respectively. 
}
\end{figure}

The main results of our theoretical calculations concerning the static 
response properties in conjunction with some selected electronic structure properties 
of small silver clusters are collected in
Table~\ref{table2} and Fig.~\ref{fig2}. Hereafter, the reported results are only for the ground state
structures. We observe that the calculated polarizabilities are in good
agreement with the experimental measurements reported in Ref.~\cite{fedrigo}. They approach
each other as the cluster size increases because the experimental data are less reliable as the
cluster size decreases. The polarizabilities reported by Fedrigo {\it et al.} \cite{fedrigo} 
were measured at T= 10 K whereas our calculated results are given at T= 0 K. Thus, according
to the following relation \cite{bottcher} in the low electric field limit
\begin{equation}
	\label{eq:3}
	\alpha_{\mathrm{eff}}=\bar{\alpha}+\frac{\mu^2}{3K_\mathrm{B}T}
\end{equation}
for clusters having a permanent dipole moment $\mu$,
the effective measured polarizability at temperature T is expected to be greater than average 
polarizability. However we have found that the contribution of the second term in
Eq.~(\ref{eq:3})--calculated at T= 10 K and for the dipole moments collected in 
Table~\ref{table2}--is negligible and consequently the influence of the temperature in 
the confrontation of the calculated polarizabilities and the experimental results
can be considered of less importance.
It is worth to note that the theoretical polarizabilities oscillate and manifest a
decreasing trend such as the experimental values do.
It converts the experimental setup designed by Fedrigo {\it et al.} 
in a valuable technique to study the electronic properties of small silver clusters.
\begin{table}
\caption{\label{table2} Calculated static response and electronic structure properties
of the lowest-energy DFT-optimized Ag$_n$ clusters.
The disproportionation energy is denoted by $\Delta_2 E_n$  and 
$\Delta\xi$ stands for the HOMO-LUMO gap.
The mean static polarizability per atom $\bar{\alpha}_{at}$ and
the polarizability anisotropy per atom $\Delta\alpha_{at}$ were calculated under
the influence of an external electric field of strength 0.0005 a.u..
The absolute value of the dipole moment is denoted by $\mu$.} 
\begin{ruledtabular}  
	\begin{tabular}{ccccccc}
		cluster & Symmetry & $\Delta_2 E_n$ & $\Delta\xi$ & $\bar{\alpha}_{at}$ & $\Delta\alpha_{at}$ & $\mu$\\
 & & (eV) & (eV) & (\AA$^3/$atom) & (\AA$^3/$atom) & (D)\\
\hline
 1\_0 &                &       & 1.24 & 6.85 & 0    & 0.06 \\
 2\_0 & D$_{\infty h}$ & 0.74  & 2.08 & 6.88 & 6.33 & 0.18\\
 3\_0 & C$_{2v}$       & -0.97 & 0.70 & 7.86 & 7.21 & 0.58\\
 4\_0 & D$_{2h}$       & 0.28  & 0.89 & 7.15 & 7.83 & 0.13\\
 5\_0 & C$_{2v}$       & -0.56 & 0.55 & 7.35 & 6.74 & 0.17\\
 6\_0 & D$_{3h}$       & 0.43  & 2.19 & 7.15 & 5.69 & 0.17\\
 7\_0 & D$_{5h}$       & -0.42 & 0.43 & 6.47 & 2.07 & 0.30\\
 8\_0 & D$_{2d}$       & 0.79  & 1.72 & 6.32 & 1.22 & 0.46\\
 9\_0 & C$_s$          & -0.73 & 0.37 & 6.62 & 3.49 & 0.54\\
 10\_0& D$_{2d}$       & 0.43  & 0.97 & 6.54 & 3.99 & 0.77\\
 11\_0& C$_1$          & -0.47 & 0.28 & 6.51 & 4.05 & 0.41\\
 12\_0& C$_s$          & 1.21  & 0.83 & 6.41 & 3.73 & 0.60\\
 13\_0& I$_h$          & -0.73 & 0.62 & 5.96 & 0.02 & 0.66\\
 14\_0& C$_{3v}$       & -0.16 & 0.43 & 5.75 & 1.25 & 0.76\\
 15\_0& C$_{2v}$       & -0.02 & 0.26 & 5.69 & 1.56 & 1.20\\
 16\_0& C$_s$          & -0.05 & 0.21 & 5.62 & 1.53 & 1.18\\
 17\_0& C$_2$          & 0.17  & 0.20 & 5.56 & 1.51 & 1.32\\
 18\_0& C$_s$          & 1.11  & 0.65 & 5.53 & 1.66 & 0.94\\
 19\_0& D$_{5h}$       & -1.10 & 0.12 & 5.67 & 2.22 & 1.15\\
 20\_0& C$_{2v}$       & 0.33  & 0.09 & 5.73 & 1.66 & 1.16\\
 21\_0& C$_1$          & 0.21  & 0.15 & 5.73 & 1.45 & 0.94\\
 22\_0& C$_s$          & 0.12  & 0.14 & 5.62 & 1.80 & 1.93\\
 23\_0& D$_{3h}$       &       & 0.19 & 5.84 & 3.06 & 1.52\\
\end{tabular}  
\end{ruledtabular}  
\end{table}
In Fig.~\ref{fig3}, we have plotted the average bond length of the lowest-energy structures 
collected in Table~\ref{table1}
 versus the cluster size. We have observed an abrupt change of the 
first-neighbor distance from Ag$_2$ to Ag$_3$ and from Ag$_6$ to Ag$_7$,
which is in our opinion a consequence of the structural transition from 1-D to 2-D and 
from 2-D to 3-D, respectively. The structural transition clearly affects the static
polarizabilities as we will see below. 
Moreover, we can see that in general the average bond length approaches to the experimental value 
of equilibrium interatomic distance of fcc silver solid (2.89~\AA) as the cluster size gets
bigger \cite{kittel}. It is a consequence of the very important role of the
surface effects in small clusters where most of the atoms belong to the
surface. 

\begin{figure}
\includegraphics[width=8.6cm,angle=0]{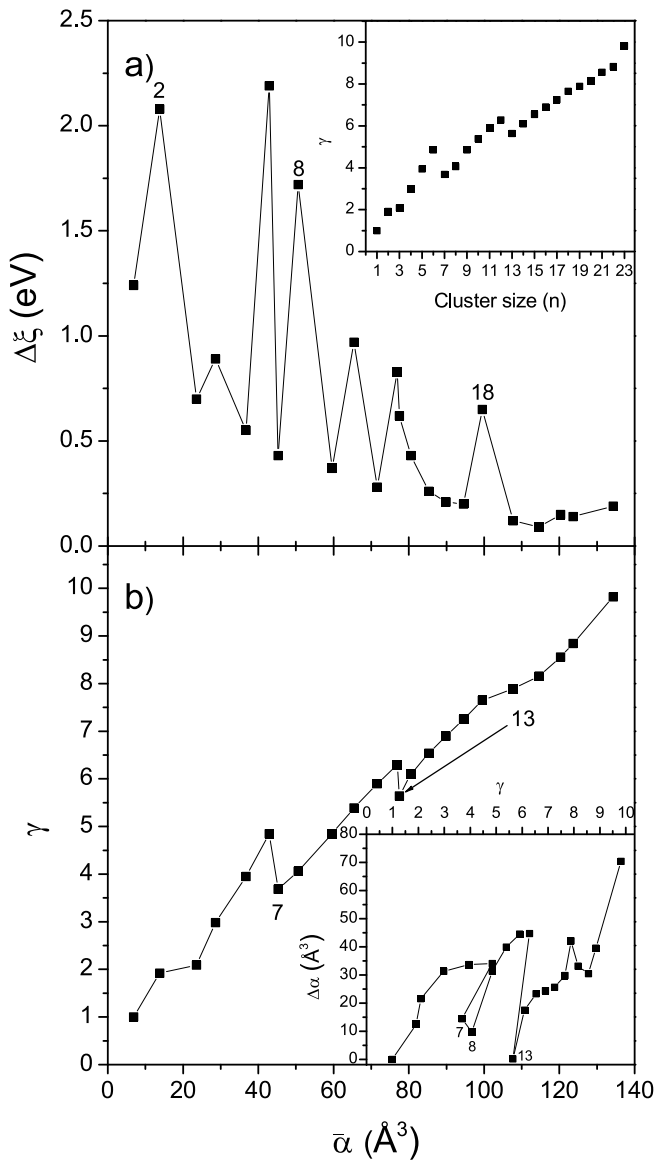}
\caption{\label{fig4}(a) The calculated HOMO-LUMO gap $\Delta\xi$ and
(b) the normalized moment of inertia $\gamma$ plotted against the mean static polarizability.
The inset in the upper panel shows the evolution of the normalized moment of inertia versus
the cluster size. In the bottom panel, the inset compares the polarizability anisotropy with
$\gamma$.}
\end{figure}

As far as the static polarizabilities displayed in Fig.~\ref{fig2} is concerned, we observe 
an odd-even oscillation of the calculated polarizability per atom in function
of the cluster size ranging from the
dimer up to the hexamer. This fact is characteristic of clusters of atoms with an odd
number of electrons and specially for atoms with a closed $d$ shell and a single valence
electron like in the case of noble-metals (Cu, Ag, and Au), 
or the closely related alkali metals (Li, Na) 
\cite{calaminici}. The even-odd oscillation of the polarizability up to the hexamer is due to
the even-odd oscillation of the HOMO-LUMO gap (see Table~\ref{table2} in conjunction with
the symmetry of the
ground state geometries). Roughly speaking an increase of the HOMO-LUMO gap is on the side of a 
chemical stabilization of the cluster but the chemical stability is favored by three-dimensional
spherical structures 
or highly symmetric two-dimensional geometries which lead to a lost of the static 
polarizability. It is
clearly reflected in Fig.~\ref{fig4}(a), where clusters with small HOMO-LUMO gaps have larger
polarizability than those with large gaps. Thus,
the static polarizability oscillates inversely as HOMO-LUMO gap does. The odd-even oscillating
trend is broken
at n=7 because of the shape transition from the planar to the compact three-dimensional structures
and is reflected by a significant decrease in the polarization of Ag$_7$ despite the fact that
the HOMO-LUMO gap decrease in relation to Ag$_6$ and Ag$_8$. In this case, the symmetry of 
the structure dominates over the HOMO-LUMO gap. 

From n=7 up to n=18, the static polarizability 
decreases smoothly in accordance with the expectation that the polarizability per atom of a cluster
is a quantity that decreases as the cluster becomes more compact and symmetric. It is borne out
by the results displayed in Fig.~\ref{fig4}(b) and the inset of Fig.~\ref{fig4}(a). With 
this purpose, we have defined the parameter
$\gamma$ that characterizes the geometry as
\begin{equation}
	\gamma=\frac{5 \mathrm{tr}\mathbf{I}}{6 n a^2}
\end{equation}
where $\mathrm{tr}\mathbf{I}$ is the trace of the  moment of inertia tensor of the 
clusters relative to the principal axis
frame, n is the number of atoms of the cluster and a stands for the radius of the silver atom
(a$\approx1.45$~\AA). In Fig.~\ref{fig4}(b) is shown that the larger clusters tend to have
higher $\gamma$ because they are structurally more elongated and consequently it is 
consistent with the classical picture that the more spherically symmetric the cluster is,
the less polarizable is. An exception is found for Ag$_7$ and Ag$_{13}$ but the reason will be 
commented below in the case of Ag$_{13}$ because 
for Ag$_7$ the structure determines the 
reduction of the polarizability as stated above. Despite the fact that the geometry is a fundamental parameter to describe
the polarizability, however it is not the only one.
It is necessary to take into account the influence of the 
HOMO-LUMO gap. Generally, the polarizability is a result of the competition between the former 
and the latter contributions, as is shown in Fig.~\ref{fig4}. Two structures 
(Ag$_8$ and Ag$_{13}$) manifest a significant reduction of the polarizability that is 
clearly reflected 
in the polarizability anisotropy which measures the symmetry or more specifically the deformation 
of the charge distribution under the influence of an external electric field in
such a way that the less the polarizability anisotropy is, the more spherically symmetric
the charge distribution is (see the inset of Fig.~\ref{fig4}(b)).  Thus, the polarizability anisotropy for Ag$_8$ and Ag$_{13}$ is
clearly reduced  since that Ag$_8$ is a closed-shell cluster with a large HOMO-LUMO gap and
Ag$_{13}$ condensates in a highly symmetric structure, i.e. the icosahedral structure (I$_h$). The 
transition from Ag$_8$ to Ag$_9$ is accompanied by an enhancement of the polarizability. It is
caused by the level structure since that whenever a new level starts to fill, the large
spatial extent of the new wave function contributes to the enhancement of the 
polarizability \cite{puska}. That is the reason for the reduced value of the polarizability in 
the case of the closed-shell structures with n=2,8, and 18.

\begin{figure}
\includegraphics[width=8.6cm,angle=0]{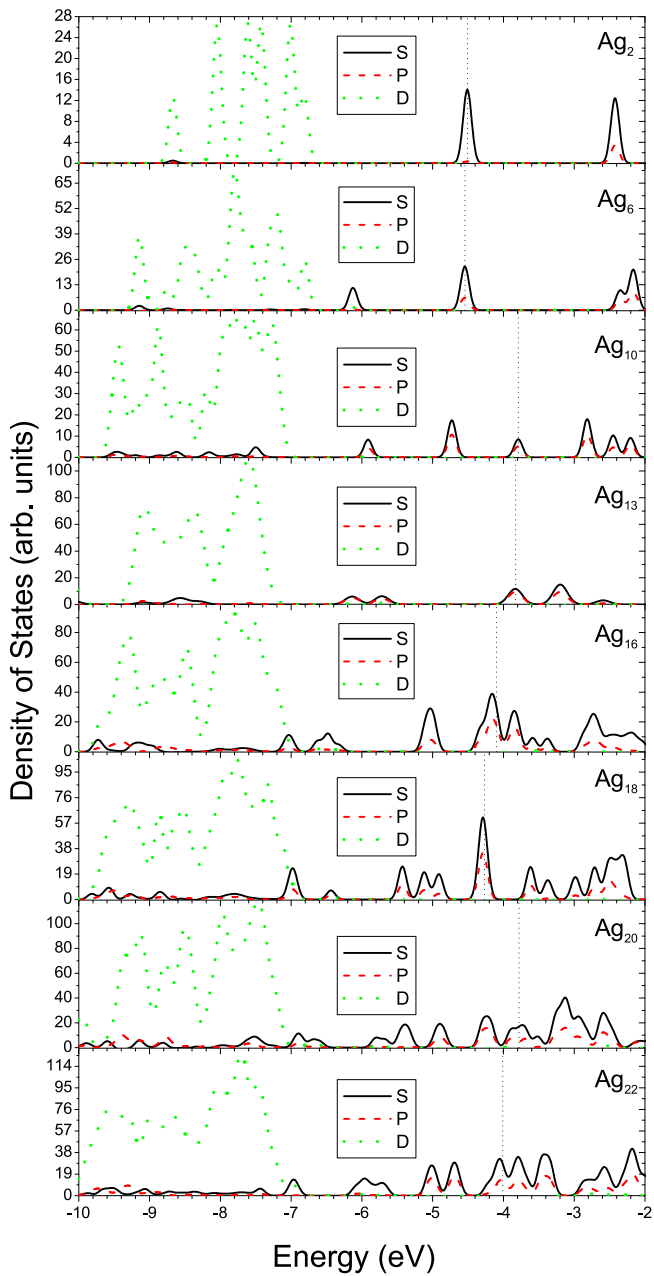}
\caption{\label{fig5}(Color online). Evolution of the partial density of states 
of the the higher-lying occupied and
lower-lying unoccupied levels with the cluster size. The solid, dashed, and 
dotted lines represent the contribution of the 5$s$, 5$p$, and 4$d$ orbitals
to the total density of states, respectively.
The dotted vertical line represents  
the Fermi level.}
\end{figure}

From n=19 up to n=23, the mean static polarizability exhibits a trend with a small positive
slope. It has been already found in Ref.~\cite{fedrigo}. Fedrigo {\it et al.} speculates that
this tendency is due to the ever increasing role of the $d$ electrons as the cluster size
grows. They state that a shift to red of the plasmon resonance due to $d$ interband transitions
corresponds to an enhancement of the polarizability. Our DFT calculations confirm this 
argument and the very important role of the $d$ electrons as the cluster increases in size.
In Fig.~\ref{fig5}, we show the evolution of the partial density of states with the cluster size.
Both $d$ and $sp$ levels gradually broaden and overlap with each other approximating to an 
electronic band in the bulk limit. With the aim of clarify the important role of $d$ electrons
with the size evolution, we have defined the energy separation $\Delta_{sd}$ as the difference
between highest
occupied molecular orbitals belonging to 4$d$ states and the lowest occupied molecular orbitals
from 5$s$ states. The energy separation decreases rapidly from 1.94 eV for Ag$_2$ to 0.31 eV for
Ag$_6$. The structural transition from planar geometry to a three-dimensional structure gives rise to 
an increase of
the value of $\Delta_{sd}$ up to 2.15 eV for Ag$_7$. After that, it decreases very rapidly
up to 0.07 eV for Ag$_{14}$. For  clusters ranging in size from n=15 up to n=23, the 
influence of 4$d$ 
level is so important that merges into the 5$s$ state. Moreover, the small HOMO-LUMO gap 
collected in Table~\ref{table2} for n=19-23 compared to smaller clusters favors the 
slightly increase of the mean static polarizability per atom as was commented above.

It is expected that the jellium model can be applied to silver clusters because the 
electronic configuration of silver is similar to the alkali clusters, where the jellium
model has been successfully applied. The solid line in Fig.~\ref{fig2} symbolizes 
the polarizability predicted by the
jellium model. We have fitted our calculated mean static polarizabilities per atom 
$\bar{\alpha}_{at}$ to an expression given by \cite{snider}
\begin{equation}
	\label{eq:5}
	\bar{\alpha}_{at}=\frac{(n^{1/3}r_{\mathrm{ws}}+\delta)^3}{n}
\end{equation}
where $r_{\mathrm{ws}}$ is the Wigner-Seitz radius and $\delta$ represents the spillout of 
the electrons from the surface of a metallic sphere. The values of the parameters resulting
from the fitting to Eq.~(\ref{eq:5}) are $r_\mathrm{ws}\approx1.63$~\AA~ and 
$\delta\approx0.45$~\AA~ which are close to the values $r_\mathrm{ws}\approx1.58$~\AA~ and 
$\delta\approx0.79$~\AA~ reported in Ref.~\cite{bennett} and Ref.~\cite{heer}, respectively.
The bulk limit of Eq.~(\ref{eq:5}) predicts a value for the bulk atomic polarizability of 
4.33~\AA$^3$/atom that is lesser than those of the the alkali metals like for example 
Na which is around 9 \AA$^3$/atom \cite{tikhonov}. As commented above, it is due to the
ever increasing role of the d electrons since that the screening of d electrons (core
polarization) tends to reduce the polarizability.
Despite that the jellium model in the spillout 
approximation predicts in average the trend of the polarizability per atom in function of the
cluster size, however it can not account for the more interesting quantum mechanical effects. Thus,
deviations of the calculated polarizabilities from the predictions of Eq.~(\ref{eq:5}) 
are ``true'' quantum 
effects. As commented above, it is in part due to the shell effects. 

In Table~\ref{table2}, we have 
collected the numerical values of the disproportionation energy, that is defined as
\begin{equation}
	\Delta_2E_n=E_{n+1}+E_{n-1}-2E_n
\end{equation}
where $E_n$ is the total energy provided by our DFT calculations of the cluster with n atoms. 
It represents the relative 
stability of a cluster with n atoms in comparison to clusters with n+1 and n-1 atoms and
consequently a peak in $\Delta_2E_n$ indicates that the cluster with size n is very stable
because a shell has been filled. The disproportionation energy shows that clusters with
n=2,8, and 18 have particularly stable configurations and consequently the polarizability is
considerably reduced, 
as is shown in Fig.~\ref{fig2}. Whenever a shell starts to fill, the polarizability increases
and deviates from the jellium model. As the size of the silver clusters increases, the HOMO-LUMO 
gap becomes smaller (see Table~\ref{table2}) so that the shell effects are less important and
the deviation of the jellium model is negligible.
\section{SUMMARY}
\label{section5}
In this article the structural stability along with the static response properties of silver clusters in the
size range $1\le n\le 23$ have been studied by means of the finite field method implemented
in the Kohn-Sham density-functional methodology \cite{salahub}. The IPs reported in this
article for the lowest-energy structures are in general in a relatively good agreement with the experimental
measurements and most of the structures predicted in this article as the fundamental ones were already reported in
former publications.
Likewise, the calculated polarizabilities are in good agreement with
the experimental measurements reported in Ref.~\cite{fedrigo}. The competition between the 
HOMO-LUMO gap and the structural symmetry on one side or the shell structure and the 
disproportionation energy on the other side are the quantum-mechanical effects that deviates
the calculated polarizabilities from the jellium model. For bigger cluster sizes the 
quantum-mechanical effects can be considered of less importance, and therefore both
theoretical approaches, i.e. the {\it ab initio} DFT calculations and the jellium model 
approach each other.

\appendix*
\section{Hellmann-Feynman theorem}
The purpose of this appendix is to show that the finite basis set 
GGA calculation holds the Hellmann-Feynman 
theorem when fully converged in the framework of density functional theory \cite{yarkony}.
We have restricted ourselves for brevity of the formulas 
to wave functions without spin polarization,
however this is not a substantial restriction and the extension to spin-unrestricted 
orbitals is straightforward.

We have selected the ansatz in which the Kohn-Sham orbitals $\psi_i(\mathrm{r})$ 
are represented by linear combinations of atomic Gaussian-type orbitals $\chi_j(\mathrm{r})$.
Thus, the orthonormal Kohn-Sham orbitals are given by:
\begin{equation}
	\label{eq:a1}
	\psi_i(\mathrm{r})=\sum_j c_{ij} \chi_j(\mathrm{r})
\end{equation}
where $c_{ij}$ are the corresponding molecular orbital coefficients. With this expansion 
we find the following relation for the electronic density:
\begin{equation}
	\label{eq:a2} 
	\rho(\mathrm{r})=\sum_{i,j} P_{ij} \chi_i(\mathrm{r}) \chi_j(\mathrm{r})
\end{equation}
where $P_{ij}$ represents an element of the density matrix, defined as 
$P_{ij}=2\sum_k^{occ} c_{ik} c_{jk}$. Using the Eq.~(\ref{eq:a1}) for the LCGTO expansions of 
the Kohn-Sham orbitals subject to the orthonormalization condition and the electronic 
density described in Eq.~(\ref{eq:a2}), the variationally
minimized Kohn-Sham
SCF energy expression may be written after some manipulation as:
\begin{eqnarray}
	\label{eq:a3}
	E_{SCF}(\lambda)&=&\sum_{ij}P_{ij}H_{ij}(\lambda)+\frac{1}{2}
	\sum_{ijkl}P_{ij}P_{kl} (ij|kl)(\lambda)\nonumber\\
	&&-2\sum_{ijk}\epsilon_k(c_{ki}c_{jk}S_{ij}(\lambda)-1)\nonumber\\
	&&+E_{xc}(\rho
	(\lambda),
	\nabla \rho(\lambda))
\end{eqnarray}
where $H_{ij}$ represents the matrix elements of the core Hamiltonian and they are built from
the kinetic and electron-nuclear interaction energies. The second term represents the
Coulomb repulsion energy of the electrons and the term $E_{xc}$ is the XC
energy in the GGA. We use the notation $(ij|kl)=\int\int \psi_i(1)\psi_j(1)(1/r_{12})\psi_k(2)
\psi_l(2) d\mathrm{r}_1 d\mathrm{r}_2$, and $\lambda$ being any parameter at all which affects
the Hamiltonian of the system. The quantities $\epsilon_k$ are one-electron eigenvalues for
the occupied orbitals and S is the overlap matrix defined as $S_{ij}=\langle\chi_i|\chi_j\rangle$.

Assuming a gradient-corrected form for the XC energy $E_{xc}=\int g(\rho,
|\nabla\rho)|^2) d\mathrm{r}$ and 
upon differentiation of the energy with respect to $\lambda$ we find
\begin{eqnarray}
	\nabla_\lambda E&=&\sum_{ij}P_{ij}\nabla_\lambda H_{ij}+\frac{1}{2}\sum_{ijkl}P_{ij}
	P_{kl}\nabla_\lambda(ij|kl) \nonumber\\
	&&-2\sum_{ijk}\epsilon_k c_{ki}c_{jk}\nabla_\lambda S_{ij}+\sum_{ij} P_{ij} (
	\langle \nabla_\lambda i|E_{xc}|j\rangle \nonumber \\
	&&+ \langle i |E_{xc}|\nabla_\lambda j\rangle +
	\langle i|\frac{\partial g}{\partial \rho} \nabla_\lambda\rho|j \rangle\nonumber\\
	&&+
	2\langle i|\frac{\partial g}{\partial |\nabla\rho|^2}\nabla_\lambda(|\nabla\rho|)|j \rangle)
\end{eqnarray}
The XC contribution to the derivative of the energy involves derivatives of the 
wave function either explicitly or implicitly throughout the electronic density. Denoting $a_i$ 
as a parameter of the wave functions that can be an exponent or positions of the bases functions,
the functional derivative of the wave function can be written as 
$|\nabla_\lambda i\rangle=|\partial i/\partial a_i\rangle (da_i/\lambda)$. Thus, we can optimize
all parameters $a_i$ so that the derivative of the XC energy can be neglected
as well as the two-electron contribution of the Hamiltonian operator of Eq.~(\ref{eq:a3}). 
Considering that the overlap matrix is independent of the perturbation $\lambda$ like for
example in the case of an uniform external electric field,
the third term depending on the derivative of the overlap matrix vanishes. Consequently, 
\begin{equation}
	\nabla_\lambda E=\langle\nabla_\lambda H\rangle
\end{equation}
which means that the fully self-consistent finite basis set solutions satisfy the Hellmann-Feynman
theorem in the framework of DFT when the XC energy is approximated by the GGA
implementation. 

\begin{acknowledgments}
The authors acknowledge the Centro de Supercomutaci\'on de Galicia (CESGA) for the computing facilities.
The work was supported by the Xunta de Galicia and the Ministerio de Educaci\'on y Ciencia under
the Projects No. PGIDIT02TMT20601PR and MAT2006-10027, respectively.
\end{acknowledgments}

\begin{thebibliography}{27}
\bibitem[{foo({\natexlab{a}})}]{footnote1}
\bibinfo{note}{During the geometry optimization process, the structure consisting of five atoms forming a
pentagon and with the other atom in the center of it converged to the pentagonal pyramid. Therefore, we have
decided to strip off any information about this structure because it is the same than the pentagonal pyramid}.
\end{thebibliography}

\end{document}